\documentclass[journal]{IEEEtran}
\usepackage{algorithm}
\usepackage{algpseudocode}
\usepackage{amsfonts}
\usepackage{bm}
\usepackage{amsmath}
\usepackage{cite}
\usepackage{amssymb}
\usepackage{bm}
\usepackage{enumerate}
\usepackage{color,xcolor}
\usepackage{graphicx}
\usepackage{multirow,tabularx}
\usepackage{subfigure}
\usepackage{hyperref}

\makeatletter

\newcommand{\Rmnum}[1]{\expandafter\@slowromancap\romannumeral #1@}
\makeatother
\ifCLASSINFOpdf
\else
\fi
\begin{document}
\title{A High Throughput Pilot Allocation for M2M Communication in Crowded Massive MIMO Systems}
\author{\IEEEauthorblockN{Huimei Han, Xudong Guo, Ying Li}}



\maketitle
\begin{abstract}
  A new scheme to resolve the intra-cell pilot collision for M2M communication in crowded massive multiple-input multiple-output (MIMO) systems is proposed.
  The  proposed scheme  permits
those failed user equipments (UEs), judged by a strongest-user collision resolution (SUCR) protocol, to contend for the idle pilots, i.e., the pilots that are not selected by any UE in the initial step. This scheme is called as SUCR combined idle pilots access (SUCR-IPA).
To analyze the performance  of the SUCR-IPA scheme, we develop a  simple method to  compute the access success probability of the UEs in each random access slot (RAST). The simulation results  coincide  well with the analysis. It is also shown that, compared to the SUCR protocol, the proposed SUCR-IPA scheme increases the throughput of the system significantly, and thus decreases the number of access attempts dramatically.
\end{abstract}
\vbox{} 
\begin{IEEEkeywords}
Massive MIMO systems, pilot collision, M2M communication, pilot allocation.
\end{IEEEkeywords}
\IEEEpeerreviewmaketitle
\section{Introduction}
\IEEEPARstart{T}{HE} massive multiple-input multiple-output (MIMO) technology can theoretically  achieve extraordinary improvements  in spectral efficiency by using a large number of antennas at a base station (BS)  \cite{Bjornson}. With the assumption that the number of antennas at the BS is infinite and the fact that the number of user equipments (UEs) is much smaller than the number of antennas at the BS, the massive MIMO channel can be viewed as an orthogonal channel offering asymptotic favorable propagation, and thus small-scale fading and thermal noise can be ignored \cite{favorable}. With these excellent properties, massive MIMO is regarded as  a key technology for time-division duplex (TDD) communication system, where downlink channel state information (CSI) can be acquired from the uplink channel estimation  through exploiting channel reciprocity \cite{Marzetta}.

In the conventional massive MIMO systems, since the number of  UEs in a cell is small,  each UE can be allocated a specialized pilot, and hence no intra-cell pilot collision occurs.
  However, such dedicated pilot allocation becomes infeasible in the fifth-generation (5G), which might contain massive number of machine-to-machine (M2M) UEs  \cite{5G}.  Therefore, pilot random access becomes a nature choice for pilot allocation  in 5G \cite{MIslam}. Under this pilot random access mechanism, the intra-cell pilot collision becomes unavoidable.
To solve this issue, J. H. S\o rensen viewed the collided pilots as a graph code and thus employed belief propagation algorithm
to alleviate pilot collision at the cost of excessive access success delays \cite{JHSrensen}.
Another interesting protocol, called as strongest-user collision resolution (SUCR) protocol, selects the UE with the strongest channel gain as the contention winner in a distributed form  \cite{Popovski}. This protocol can improve the probability of collision resolution under low delay.
However, as the authors pointed out, the SUCR  protocol always regards the strongest one as the winner, which is unfair for the weaker UEs. Our further research show  that the collision resolution probability of the SUCR protocol is decreasing with the increase of the number of contending UEs.
Specifically, the more failed UEs in current random access slot (RAST), the more contenders in their related RASTs during which the failed UEs will reattempt their accesses.  As a result,  with the increase of the failed UEs  in current RAST, the number of failed UEs in its related RASTs will increase.

   To conquer this issue, we propose a new scheme which can further  improve the throughput, namely the number of UEs who are successfully allocated  pilots. The proposed scheme permits those failed UEs, judged by the SUCR protocol, to contend for the idle pilots. Hence, the throughput increases and the fairness between UEs can be ensured to a certain extent. We call this scheme as SUCR combined idle pilots access (SUCR-IPA).
Employing the system model of  random access procedure proposed in \cite{model}, we establish a  simple method to compute
the access success probability of UEs in each RAST,  or the failed probability. Based on this method, we further   analyze  the performance of the proposed SUCR-IPA scheme, including the throughput during a certain RAST,  the access success probability during the observed RASTs and the cumulative density function (CDF) of
   the number of access attempts. Simulation results show that,  compared to the SUCR protocol, the throughput of the SUCR-IPA scheme increases significantly and the average  number of access attempts decreases dramatically. Finally, simulation results are provided to verify the validity of the analysis results.

The remainder of this paper is organized as follows. System model and the principle of SUCR-IPA scheme are described  in Section II. Section III elaborates the performance analysis  of the SUCR-IPA scheme. Simulation results and the conclusion are presented in Section IV and V, respectively.
\section{System Model and The Proposed SUCR-IPA  Scheme}
\subsection{System model}

We consider a single BS equipped with $M$ antennas at the center of a hexagonal network in TDD MIMO communication system, and there are $K$ single-antenna UEs in this system.

\begin{figure}
  \centering
  \includegraphics[width=9cm,height=5cm]{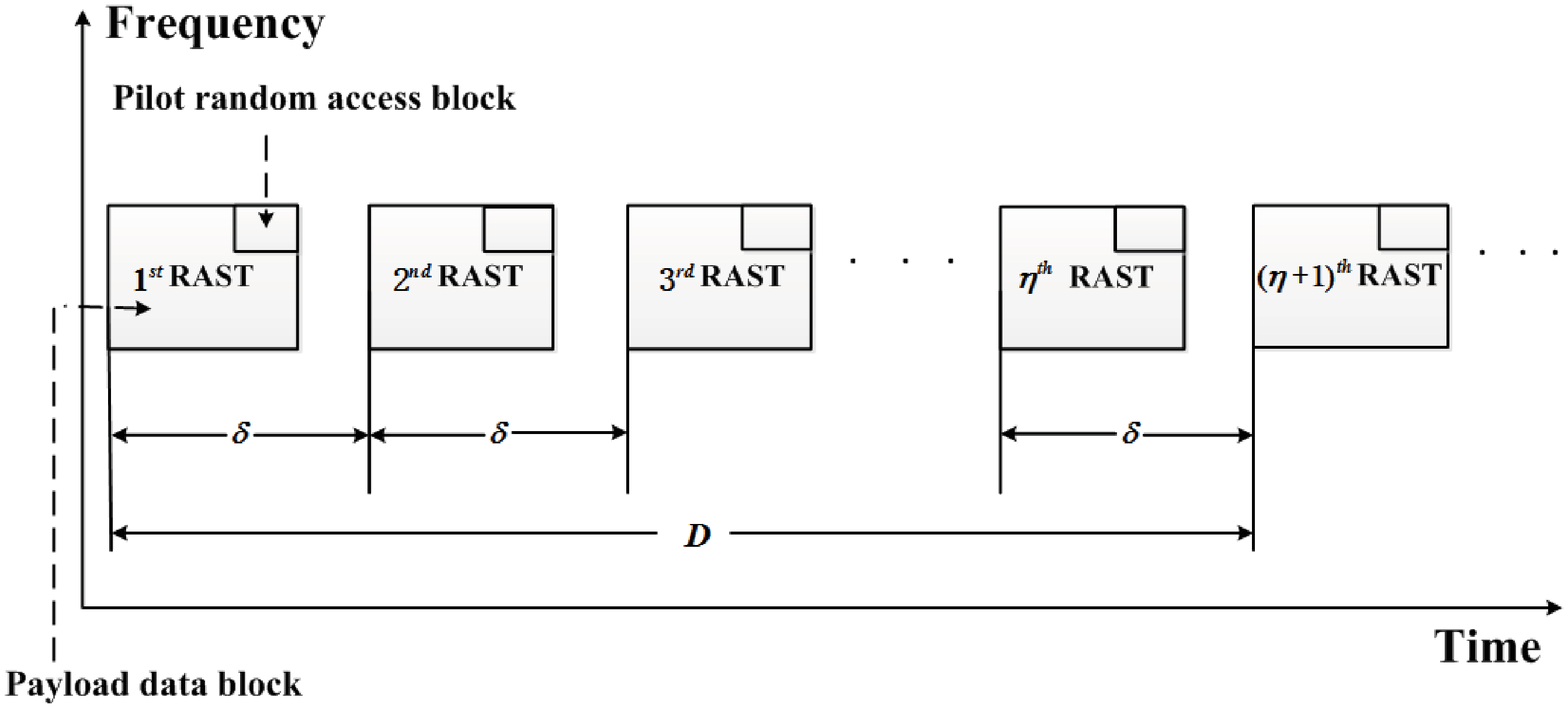}\\
  \caption{System model.}\label{System}
\end{figure}

In this paper, we consider the pilot random access procedure which is performed in time slot. As illustrated in Fig.\ref{System},
in each RAST, the total time-frequency resource is divided into two blocks, namely pilot random access block and payload data block. As their names shown, the pilot random access block is used by UEs to access randomly to the pilot, and the payload data block is used to transmit the UEs' payload data, which is the same as described in \cite{Popovski}.  Only when the UE accesses to the pilot successfully, can it send its payload data. In this paper we only focus on the pilot random access procedure.

Assume that the interval between any two successive RASTs is $\delta$. The ${{i}^{th}}$ RAST, during the observed time $[0,{D}]$, is denoted by $\text{R}{{\text{S}}_{i}}$, $1\le i\le \eta$ .
Let $Z_{i}^{n}$  denote the number of new arrivals performing their ${{n}^{th}} $ $(1\le n\le W)$  access attempts during $\text{R}{{\text{S}}_{i}}$, where  $ W$  is the maximum number of access attempts. Thus, the number of active UEs during $\text{R}{{\text{S}}_{i}}$  can be written as
  \begin{equation}\label{1}
    {{Z}_{i}}=\sum\limits_{n=1}^{W}{Z_{i}^{n}}.
  \end{equation}
According to \cite{3GPP},  $Z_{i}^{1}$  is defined as
\begin{equation}\label{2}
 {Z_{i}^{1}}=N\int\limits_{{{t}_{i}}}{g(t)}dt,
\end{equation}
where $N$ is the total number of new arrivals during $\eta$ RASTs, ${{t}_{i}}=\delta \times i$, and  ${g(t)}$ is the probability density function of M2M calls. As described in  \cite{3GPP}, ${g(t)}$ is defined as
\begin{equation}\label{3}
 g(t)=\frac{{{t}^{\alpha -1}}{{(\text{T-}t)}^{\vartheta -1}}}{{{\text{T}}^{\alpha +\vartheta -1}}\beta (\alpha ,\vartheta )},\text{                          0}\le t\le D ,
\end{equation}
 where $\beta (\alpha ,\vartheta )$ is the beta function with parameters $\alpha $ and $\vartheta $, and  defined as
 \begin{equation}\label{27}
   \beta (\alpha ,\vartheta ) = \int\limits_0^1 {{t^{\alpha  - 1}}{{(1 - t)}^{\vartheta  - 1}}}.
 \end{equation}
\subsection{The Proposed SUCR-IPA scheme}
  \begin{figure}
  \centering
  \includegraphics[width=7cm,height=7cm ]{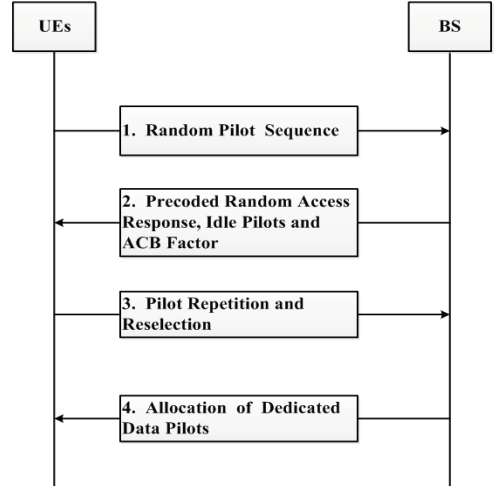}\\
  \caption{The proposed SUCR-IPA scheme.}\label{scheme}
\end{figure}
Fig.\ref{scheme} shows the main four steps of the SUCR-IPA scheme, whose main idea is to make full use of the idle pilots.
Consider the ${{i}^{th}}$ RAST, $\text{R}{{\text{S}}_{i}}$,  $1\le i\le \eta$.  Assume that the  number of active UEs, ${{Z}_{i}}$, is available to the BS. The details of this scheme are described as follows.

Step 1:  UE Randomly Selecting Pilot Sequence

 Each active M2M UE randomly chooses a pilot from the set of mutually orthogonal pilots $\bm{P_o}=\left\{ \bm{{\xi }_{1}},\bm{{\xi }_{2}},\bm{{\xi }_{3}}\cdots \bm{{\xi }_{{{\tau }_{p}}}} \right\}$ with equal probability $\frac{1}{{{\tau }_{p}}}$ and transmits it to the BS.

Let $\bm{h_k}={(h_k^1,h_k^2, \cdots ,h_k^M)^{\text{T}}}$ denote the channel gain between UE $k$ and the BS, where $(\centerdot)^{\mathrm{T}}$ is the transpose operation. Let $\bm{{\psi }_{k}}$ be the pilot sequence selected by UE $k$ and satisfies  ${||\bm{{\psi }_{k}}||}=\sqrt{{{L }}}$, where $L$ is the length of each pilot and $|| \bullet ||$ stands for the Euclidean norm of a vector. Let ${{\rho }_{k}}$ denote the uplink  transmitting power of UE $k$.
The received pilot signal ${\bm{Y}}$ at BS is
\begin{equation}\label{10}
  {\bm{Y}}=\sum\limits_{k=1}^{{{Z}_{i}}}{\sqrt{{{\rho }_{k}}}\text{ }{\bm{{h}_{k}}}\bm{\psi _{k}}^{\mathrm{T}}+\bm{N}},
\end{equation}
where  $\bm{N}\in {{\mathbb{C}}^{\text{  }M\times {L}}}$ is white noise distributed vector (or matrix) with each element being mean zero and variance ${{\sigma }^{2}}$  circularly-symmetric complex Gaussian distribution, i.e.,  $ CN(0,{{\sigma }^{2}})$, and $\mathbb{C}$ is the space of complex-valued.

In this paper, we  consider the uncorrelated Rayleigh fading channels,  i.e.,
$\bm{{h}_{k}}\sim \text{ }CN(0,{{\beta }_{k}}\times{\bm{{{I}}_{M}}})$,  where $\bm{{I}_{M}}$ denotes the $M\times M$ identity matrix and ${{\beta }_{k}}$ accounts for  the path loss of UE $k$.

Step 2: BS Generating and Broadcasting Precoded Random Access Response (PRAR), Idle Pilots and
 Access  Class Barring
(ACB) Factor

To obtain the desired information,  following the procedure in \cite{SUCR}, we first correlate $\bm{Y}$ with each of the pilots in $\bm{P_o}$. Thus, we have
              \begin{equation}\label{11}
                {\bm{{{y}}_{t}}}=\bm{Y}\frac{\bm{\xi _{t}}^{*}}{||{\bm{{\xi }_{t}}}||} \ \ \ ,\text{                                                          1}\le t\le {{\tau }_{p}},
              \end{equation}
              where ${\bm{{{y}}_{t}}}$ is a vector with $M$ elements and ${(\centerdot)^{*}}$ denotes the conjugate of a vector (or matrix).


  According to the remark 1 in \cite{SUCR}, when $M$ is large, the value of $\frac{||\bm{{{y}}_{t}}||^2}{M}$  of the idle pilot, i.e., the pilot that  is not selected by any  UE, almost equals the variance of the additive noise, while that of the selected pilot almost equals the sum of the signal gains and  the variance, which is much greater than that of the idle pilot. Hence,  the BS can easily  estimate the number and indexes of those idle pilots, denoted by  ${{\text{G}}_{i}}$ and $\left\{ {{n}_{1}},{{n}_{2}},{{n}_{3}},\ldots ,{{n}_{{{G}_{i}}}} \right\}$ accordingly. The PRAR can be calculated as  \cite{SUCR}
 \begin{equation}\label{12}
                \bm{V=}\sqrt{q}\sum\limits_{t=1}^{{{\tau }_{p}}}{\frac{\bm{{y}_{t}}}{||{\bm{{y}_{t}}}||}}\bm{\bm{\phi  }_{t}}^{\mathrm{T}},
              \end{equation}
              where $q$ is the downlink transmitting power, and $\bm{\bm{\phi  }_{t}}$ is the downlink pilot corresponding to  the
              ${{t}^{th}}$ uplink pilot.

Finally, we estimate the expected number of UEs, who will not repeat their pilots during step 3, denoted by $F$. Thus,  the BS gets the ACB factor ${T_{i}^{v}}$ as \cite{ACB}
\begin{equation}\label{17}
  T_i^v = \left\{ \begin{gathered}
  \frac{{{G_i}}}{F}{\text{                             }}, F > {G_i}, \hfill \\
   \hfill \\
  1{\text{                                }}, F \leqslant {G_i}. \hfill \\
\end{gathered}  \right.
\end{equation}
The estimation of  $F$ will be described in Section III.

 After these processes, the BS broadcasts PRAR, idle Pilots indexes and ACB factor to all active UEs via downlink broadcast channel.

Step 3: UE Transmitting either Repeated  or Reselected Pilot

Based on the received PRAR, each UE independently determines whether it is the strongest UE, following the method described in \cite{Popovski}. If it is a winner, the UE will repeat its pilot to the BS. Otherwise, the UE generates a random value  ranging from 0 to 1, and compares it with the ACB factor ${T_{i}^{v}}$. When the generated random value is less than ${T_{i}^{v}}$, the UE will contend for the ${{{\text{G}}_{i}}}$ idle pilots and send the reselected pilot to the BS. If the generated random value is lager than ${T_{i}^{v}}$, the UE remains silent, which implies that the UE is failed to access to the pilot under current RAST. In addition, along with the pilot sequence, UEs should also transmit uplink messages such as the identity numbers of the UEs during this step.

Step 4: BS Allocating Dedicated Data Pilots (DDP)

After receiving the pilots, the BS estimates the channel gain of each UE and utilizes it to decode the corresponding UL message. If the decoding successes, the BS allocates DDP to the corresponding UE, a procedure resembling step 2. Those UEs, who do not receive the DDP, will select an integer number ${B}$ from 1 to ${{W}_{BO}}$ uniformly. Then, after waiting ${B}$ time, the UE reattempts its access in the upcoming RAST.

\section{Performance analysis}
In this section, we mainly analyze the performance of the SUCR-IPA scheme, including the throughput of the system   during the ${{i}^{th}}$ RAST, access success probability during  ${\eta}$  RASTs, and the CDF of the number of access attempts.

The number of access success  UEs, denoted by $ Z_{i}^{s}$,  represents the throughput of the system and can be computed by
\begin{equation}\label{18}
\begin{array}{rcl}
Z_i^s & =& {\tau _p}\sum\limits_{u = 1}^{{Z_i}} {{P_r}(D_u^1)} \left( \begin{array}{l}
{Z_i}\\
u
\end{array} \right){\left( {\frac{1}{{{\tau _p}}}} \right)^u}{\left( {1 - \frac{1}{{{\tau _p}}}} \right)^{{Z_i} - u}}\\
&+&{G_i}\left( \begin{array}{l}
F\\
1
\end{array} \right)\left( {\frac{{T_i^v}}{{{G_i}}}} \right){\left( {1 - \frac{{T_i^v}}{{{G_i}}}} \right)^{F - 1}}\\
&\mathop  = \limits^{(a)}& {\tau _p}\sum\limits_{u = 1}^{{Z_i}} {{P_r}(D_u^1)} \left( \begin{array}{l}
{Z_i}\\
u
\end{array} \right){\left( {\frac{1}{{{\tau _p}}}} \right)^{u - 1}}{\left( {1 - \frac{1}{{{\tau _p}}}} \right)^{{Z_i} - u}}\\
& + &F \times T_i^v \times {\left( {1 - \frac{{T_i^v}}{{{\tau _p} \times {{(1 - \frac{1}{{{\tau _p}}})}^{{Z_i}}}}}} \right)^{F - 1}},
\end{array}
\end{equation}
where  $(a)$ follows from the fact that ${{\text{G}}_{i}}={{\tau }_{p}}\times {{(1-\frac{1}{{{\tau }_{p}}})}^{{{Z}_{i}}}}$ and $Pr(D_u^1)$ denotes the probability that a pilot sequence is selected by $u$ UEs during step 1, while there is only one UE repeating this pilot during step 3. It can be seen easily that $u$, i.e., the number of M2M UEs who select the same pilot, follows a binomial distribution.  We denote it as ${{u}}\sim{}B({{{Z}_{i}}},\frac{1}{{{\tau }_{p}}})$.

 We note that $ Z_{i}^{s}$ includes two terms. The first term
 ${\tau _p}\sum\limits_{u = 1}^{{Z_i}} {{P_r}(D_u^1)\left( \begin{gathered}
  {Z_i} \hfill \\
  u \hfill \\
\end{gathered}  \right){{(\frac{1}{{{\tau _p}}})}^u}{{(1 - \frac{1}{{{\tau _p}}})}^{{Z_i} - u}}}$
indicates that the number of access success UEs who repeat their pilots during step 3. The second term $F \times T_i^v \times {\left( {1 - \frac{{T_i^v}}{{{\tau _p} \times {{(1 - \frac{1}{{{\tau _p}}})}^{{Z_i}}}}}} \right)^{F - 1}}$ denotes the number of access success UEs who select the idle pilots during step 3. Now, we discuss how to compute the  probability  ${P_r}(D_u^1)$, the expected number of UEs who do not repeat their pilots $F$ during step 3, and the ACB factor $T_i^v$.

First let ${D_{{{u}}}^{d}}$ denote the event  that a pilot is selected by $u$ UEs during step 1 while there are ${d}$ ($0\le d\le {u}$) UEs  repeating this pilot during step 3. Let  ${{{R}_{k}}}$  denote the event that UE ${k}$ is able to repeat its pilot and  ${{{J}_{k}}}$ denote the event that UE ${k}$ is not able to repeat its pilot.

For event ${D_{{{u}}}^{d}}$, it can be seen easily that there are $\lambda _{{u}}^{d}$ kinds of different cases with respect to the fact that, among u UEs, there are $d$ UEs repeat this pilot during step 3.
 For the ${{l}^{th}}$ ($1\le l\le \lambda _{{u}}^{d}$) case, we use $\bm{{{X}}_{l}}=\left\{ x_{l}^{1},x_{l}^{2},\cdots ,x_{l}^{d} \right\}$ to denote the indexes of $d$ UEs who repeat this pilot, and the remaining $u-d$ UEs are denoted by $\bm{{{Y}}_{l}}=\left\{ y_{l}^{1},y_{l}^{2},\cdots ,y_{l}^{u-d} \right\}$. The probability of ${D_{{{u}}}^{d}}$ can be written as
 \begin{equation}\label{13}
               {{P}_{r}}(D_{{{u}}}^{d})=\sum\limits_{l=1}^{\lambda _{{{u}}}^{d}}{{{P}_{r}}\left\{ {{R}_{x_{l}^{1}}},\cdots ,{{R}_{x_{l}^{d}}},{{J}_{y_{l}^{1}}},\cdots ,{{J}_{y_{l}^{u-d}}} \right\}}.
              \end{equation}

By setting $d$ in (\ref{13}) to 1, we get the probability  ${P_r}(D_u^1)$.

The computation of  ${{{P}_{r}}\left\{ {{R}_{x_{l}^{1}}},\cdots ,{{R}_{x_{l}^{d}}},{{J}_{y_{l}^{1}}},\cdots ,{{J}_{y_{l}^{u-d}}} \right\}}$  in (\ref{13}) with respect to $ d=1$ can be found in \cite{SUCR}, and the same way can be used to  get the value of this term corresponding to other values of $d$.

The expected number of UEs who do not repeat their pilots during step 3,  can be calculated by
\begin{equation}\label{16}
 F=\sum\limits_{u=1}^{{{Z}_{i}}}{{{F}_{u}}},
\end{equation}
where ${{F}_{u}}$,  the expected number of UEs  not  repeating their pilots  during step 3 among ${{u}}$ contenders, can be calculated by
\begin{equation}\label{15}
  {{F}_{{{u}}}}=\sum\limits_{d=0}^{{{u}}}{({{u}}-d}){{P}_{r}}(D_{{{u}}}^{d})C_{{{\tau }_{p}}}^{{{u}}},
\end{equation}
where $C_{{\tau _p}}^u$,  the expected number of pilots selected by ${{{u}}}$ UEs, can be computed by
\begin{equation}\label{14}
\begin{array}{rcl}
C_{{\tau _p}}^u &=& \sum\limits_{m = 1}^{{\tau _p}} {} \left( \begin{array}{l}
{{Z_i}}\\
u
\end{array} \right){(\frac{1}{{{\tau _p}}})^u}{(1 - \frac{1}{{{\tau _p}}})^{{Z_i} - u}}\\
&{\rm{               = }}&{\tau _p}\left( \begin{array}{l}
{{Z_i}}\\
u
\end{array} \right){(\frac{1}{{{\tau _p}}})^u}{(1 - \frac{1}{{{\tau _p}}})^{{Z_i} - u}},
\end{array}
\end{equation}

Let ${{P}_{s}} $ denote the access success probability  during $\eta$ RASTs. Apparently, ${{P}_{s}} $ is just  the ratio of the throughput to the number of active UEs  during $\eta$ RASTs. Therefore, we have
  \begin{equation}\label{7}
    {{P}_{s}}=\frac{\sum\limits_{i=1}^{\eta }{{}}{Z_{i}^{s}}}{\sum\limits_{i=1}^{\eta }{{{Z}_{i}}}}.
  \end{equation}

Let ${{F}_{p}} $ represent the CDF of  the number of access attempts. Then, we have
  \begin{equation}\label{8}
    {{F}_{p}}(p\le r)=\frac{\sum\limits_{i=1}^{\eta }{{}}\sum\limits_{n=1}^{r}{Z_{i,s}^{n}}}{\sum\limits_{i=1}^{\eta }{{}}\sum\limits_{n=1}^{W}{Z_{i,s}^{n}}},
  \end{equation}
where ${Z_{i,s}^{n}}$ is the number of UEs, who are successfully allocated pilots  after  ${{n}} $  access attempts during $\text{R}{{\text{S}}_{i}}$. Let $P_{UE,i}^{s}$ represent the access success probability of UEs during $\text{R}{{\text{S}}_{i}}$. Hence, we have
 \begin{equation}\label{6}
    Z_{i,s}^{n}={{Z}_{i}^{n}}\times P_{UE,i}^{s}.
  \end{equation}

 We assume that the number of active UEs  during the ${{i}^{th}} $ RAST, i.e.,  ${{Z}_{i}}$, is known. Furthermore,  the throughput during the ${{i}^{th}}$ RAST, $ Z_{i}^{s}$, which depends on the access scheme, can be derived easily by (\ref{18}). Hence, $P_{UE,i}^{s}$ can be calculated by
 \begin{equation}\label{5}
   P_{UE,i}^{s}=\frac{Z_{i}^{s}}{{{Z}_{i}}}.
 \end{equation}

Apparently,  the failed probability of UEs during $\text{R}{{\text{S}}_{i}}$ is $P_{UE,i}^{f}=1-P_{UE,i}^{s}$.

It should be noted that the calculation of $P_{UE,i}^{s}$ in (\ref{5}) is different from that mentioned in \cite{model}. The main advantage of (\ref{5}) is that the calculation is suitable for almost all the random access schemes, since the value of $ Z_{i}^{s}$  can  be obtained either by analysis or by simulation.   The computation of $P_{UE,i}^{s}$ in \cite{model} is given in a direct way. However, it is hard to obtain the value of  $P_{UE,i}^{s}$ directly for the SUCR protocol proposed in \cite{Popovski}.


\section{Simulation Results}
In this section, we compare the performance of the SUCR-IPA scheme with  SUCR protocol, in terms of the throughput during a certain RAST, the access success probability during  $\eta$ RASTs  and the CDF of the number of access attempts.

 The path loss exponent of the uncorrelated Rayleigh fading in the urban micro scenario is 3.8 \cite{Spatial}.
 We assume that the  UEs and BS in the cell transmit signals at full power, i.e., ${{\rho }_{k}}=q=1$. The median signal to noise ratio (SNR) of the UEs at the corner of cell is 0 dB. The radius of the cell is 250 meters and all UEs locate uniformly at the place which is farther than 25 meters from the BS.

Consider the crowded scenario that there are $N$ active M2M UEs during $\eta$ RASTs with interval $\delta$, where $N=2000$, $\eta  =100$, and $\delta =10ms$.
 We set the number of pilots allocated to M2M UEs as ${{\tau }_{p}}=60$ and the length of each pilot is $L=64$. We also set ${{W}_{BO}}=20ms$ and ${{W}}=10$.
\begin{figure}
  \centering
  \includegraphics[width=9cm,height=6cm]{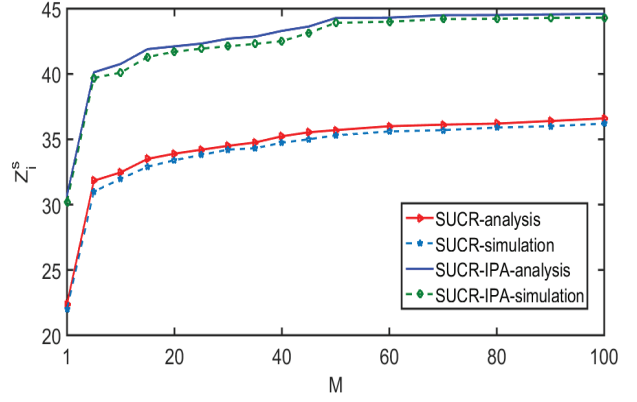}\\
  \caption{The throughput ${{Z}_{i}^{s}}$  during the $i^{th}$ RAST.}\label{successful}
\end{figure}

Fig.\ref{successful} illustrates the variance of the throughput ${{Z}_{i}^{s}}$ with $M$ antennas.   The number of active UEs in $\text{R}{{\text{S}}_{i}}$ is set as ${{Z}_{i}}=60$. This scenario indicates that the system is fundamentally overloaded in the sense that, on average, each pilot is selected by one UE. Both simulation  results and analysis results are included in this figure. Another point should be noted is that the analysis results of the SUCR-IPA scheme and the SUCR protocol can be obtained by (\ref{18}) and the first term of (\ref{18}), respectively. We can see that the simulation results match well with the analysis results. Furthermore, the throughput of the SUCR-IPA scheme is significantly higher  than that of the SUCR protocol. We can also note that the throughput ${{Z}_{i}^{s}}$  increases dramatically from ${M}=1$ to ${M}=20$, and increases at a slower pace when ${M}\ge20$.
\begin{figure}
  \centering
  \includegraphics[width=9cm, height=6cm]{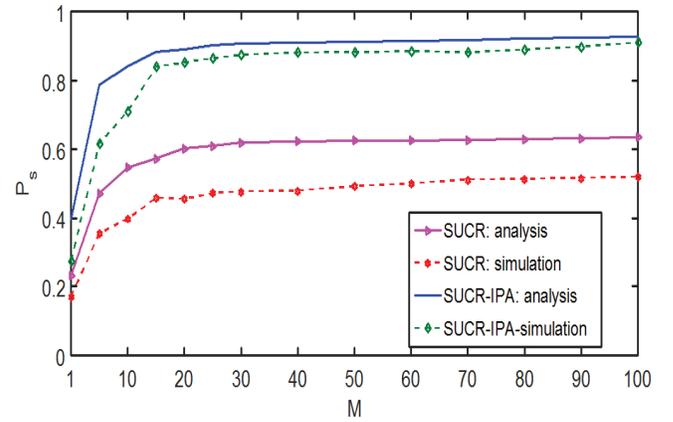}\\
  \caption{Access success probability ${{P}_{s}} $ during  $\eta$ RASTs where $\eta=100$.}\label{probability}
\end{figure}

\begin{figure}
  \centering
  \includegraphics[width=9cm,height=6cm]{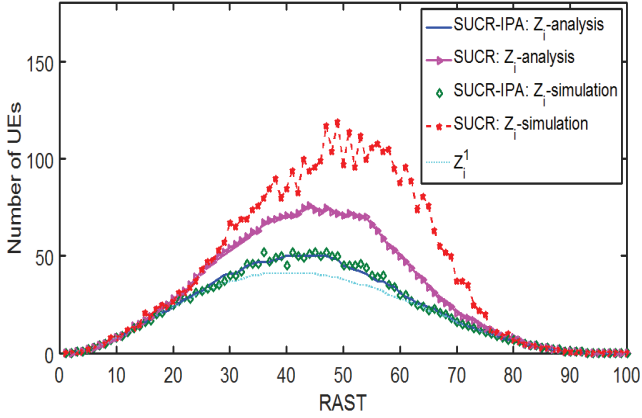}\\
  \caption{The  number of active UEs ${{Z}_{i}}$ and new arrivals ${{Z}_{i}^{1}}$ during the ${{i}^{th}}$ RAST for $1\le i\le \eta$, where $\eta =100$.}\label{totalnumber}
\end{figure}

Fig.\ref{probability} shows the access success probability ${{P}_{s}} $ during the $\eta$  RASTs, where $\eta=100$. We can observe that the  simulation results of the SUCR-IPA scheme match well with our analysis results when ${M}\ge20$ and the ${{P}_{s}} $ is as high as 90$\%$. In contrast,  ${{P}_{s}} $  of the SUCR protocol is much smaller than that of the SUCR-IPA scheme and the simulation results does not match well with its analysis results. The reason for the results can be explained from the impact of the current RAST on its related RASTs.  Fig.\ref{totalnumber}  gives
the number of active UEs in (\ref{1}) and new arrivals in (\ref{2}) for each RAST when ${M}=50$.
Due to the high  access success probability of  UEs in each RAST of  the SUCR-IPA scheme,  the number of UEs in the current RAST who reattempt accesses in its related RASTs is small. Hence,
 the increased number of UEs in its related RASTs caused by the failed UEs in the current RAST is very small. In other words, the impact of the current RAST on its related RASTs will be small. Therefore, we can note that the  number of active UEs is close to the number of new arrivals, as shown in Fig.\ref{totalnumber}. Furthermore, recalling conclusion in \cite{SUCR} that, the more contenders in the RAST, the smaller the probability of only one UE among the contenders repeating its pilot during step 3, we observe that the current RAST will almost not impact the number of failed UEs in its related RASTs and not further impact the number of access attempts.
 Nevertheless, for the SUCR protocol, due to the lower access success probability of  UEs in each RAST compared to the SUCR-IPA scheme, the impact of current RAST on its related RASTs as described above is large, and hence the  number of active UEs is far away from the number of new arrivals as shown in Fig.\ref{totalnumber}. As a result, the current RAST greatly impact the number of failed UEs in its related RASTs.
\begin{figure}
  \centering
  \includegraphics[width=9cm,height=6cm]{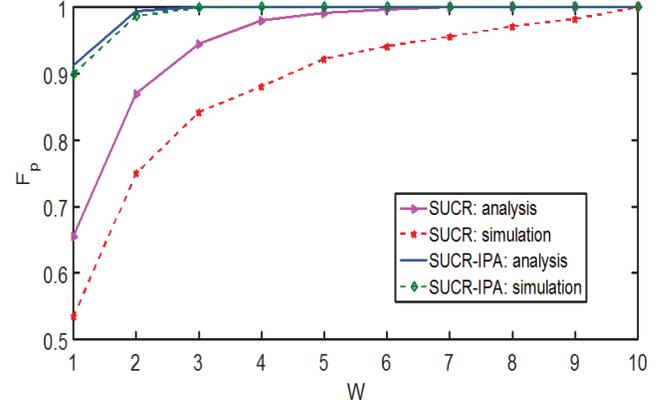}\\
  \caption{CDF of number of access attempts  ${{F}_{p}}$.}\label{transmissionnumber}
\end{figure}

Fig.\ref{transmissionnumber} depicts the CDF of the number of access attempts ${{F}_{p}} $  when ${M}=50$. We can note that the simulation results of the SUCR-IPA scheme is almost identical to its analysis results, and almost 90$\%$  UEs are successfully allocated pilots  in exactly one access attempt. However, for the SUCR protocol, only 55$\%$ UEs are successfully allocated pilots  in one access attempt. The reason for this is similar to that we explained for Fig.4. That is, the current RAST in the SUCR-IPA scheme  has almost no effect on  the number of failed UEs in its related RASTs, while that in the SUCR scheme has great effect.
\section{Conclusion}
In this paper, we propose a new scheme for M2M communication in crowded massive MIMO systems to resolve the intra-cell pilot collision. The idle pilots  provide  another opportunity for those failed UEs judged by SUCR protocol  to be allocated pilots successfully.
We also  propose a  simple method to compute the access success
probability for UEs per RAST. The simulation results match well with analysis results  and show that the SUCR-IPA scheme gives much better performance than the SUCR protocol.

\appendices

\end{document}